\documentclass{PoS}
\usepackage{graphicx}
\usepackage{amssymb}
\usepackage{epstopdf}
\usepackage{amsmath}
\usepackage{array}
\usepackage{color}
\usepackage{slashed} 
\usepackage{braket}

\newcommand{\LSP}{\widetilde{\chi}_1^0}
\newcommand{\neutralinotwo}{\widetilde{\chi}_2^0}
\newcommand{\charginoone}{\widetilde{\chi}_1^{\pm}}

\title{Naturalness and light Higgsinos: why ILC is the right machine for SUSY discovery}

\ShortTitle{Naturalness and light Higgsinos: why ILC is the right machine for SUSY discovery}

\author{Howard Baer \\%
        University of Oklahoma,
        Norman, OK 73019, USA \\
        E-mail: \email{baer@ou.edu}}

     \author{Mikael Berggren, \speaker{Suvi-Leena Lehtinen}\thanks{On behalf of the ILD concept group},
                Jenny List\\
     DESY, Hamburg, Germany\\
     E-mail:  \email{mikael.berggren@desy.de}, \email{suvi-leena.lehtinen@desy.de},
     \email{jenny.list@desy.de}}

\author{Keisuke Fujii, Jacqueline Yan\\
        KEK,        Tsukuba, Japan \\
        E-mail: \email{keisuke.fujii@kek.pj, jackie@post.kek.jp}}
        
\author{Tomohiko Tanabe\\
        International Center for Elementary Particle Physics, University of Tokyo,
        Tokyo, Japan
        E-mail: \email{tomohiko@icepp.s.u-tokyo.ac.jp}}        

\abstract{Radiatively-driven natural supersymmetry, a theoretically and experimentally well-motivated framework, centers around the predicted existence of four light, nearly mass-degenerate Higgsinos with mass $\sim 100-200$ GeV (not too far above $m_Z$).
The small mass splittings amongst the higgsinos, typically 4-20 GeV, results in very little visible energy arising from decays of the heavier higgsinos. Given that other SUSY particles are considerably heavy, this makes detection challenging at hadron colliders. On the other hand, the clean environment of an electron-positron collider with  $\sqrt{s}>2m_{higgsino}$ would enable a decisive search of these required higgsinos, and thus either the discovery or exclusion of natural SUSY.  We present a detailed simulation study of precision measurements of higgsino masses and production cross sections at  $\sqrt{s}$ = 500 GeV of the proposed International Linear Collider currently under consideration for construction in Japan. The study is  based on a Geant4 simulation of the International Large Detector concept. We examine several benchmark points just beyond the HL-LHC reach, with four light higgsinos directly accessible by the ILC, and the mass differences between the lightest SUSY particle and the heavier states ranging from about 4 to 20 GeV. It can be shown that their masses and production cross sections can be precisely measured to approximately 1\% precision or better. These precise measurements allow for extracting the underlying weak scale SUSY parameters, giving predictions for the masses of heavier SUSY states. These provide motivation for future high-energy colliders. Additionally, dark matter properties may be derived. Evolution of the measured gaugino masses to high energies should allow testing the hypothesis of gaugino mass unification.
}

\FullConference{The European Physical Society Conference on High Energy Physics\\
                 5-12 July 2017\\
                 Venice, Italy}

\begin{document}

\section{Motivation}

While LHC has discovered a Standard-Model-like Higgs boson with $m_h\simeq 125$ GeV, its 
existence within the Standard Model is hard to comprehend due to unstable quantum corrections to its mass.
Supersymmetry (SUSY) tames the quadratic divergences while successfully predicting the Higgs mass to lie
within the narrow range $m_h\sim 115-135$ GeV. 
However, the fact that no SUSY particles have appeared thus far at LHC resurrects the fine-tuning issue in the
form of the Little Hierarchy problem. 
The connection between the magnitude of the weak scale 
and MSSM Lagrangian
parameters arises from the scalar potential minimisation condition:
\begin{equation}
 \frac{m^2_Z}{2} = \frac{(m^2_{H_d}+ \Sigma_d^d) - (m^2_{H_u}+\Sigma_u^u)\tan^2 \beta}{\tan^2\beta -1}-\mu^2
\simeq -m_{H_u}^2-\Sigma_u^u-\mu^2.
\end{equation}
To avoid fine-tuning, thus maintaining naturalness, $m_{H_u}^2$ should be driven to small negative values, 
and the superpotential $\mu$ parameter should be $\mathcal{O}(100-300)$ GeV (not too far from $m_{weak}\sim 100$\,GeV). 
The radiative corrections $\Sigma_u^u$, evaluated at the weak scale, are minimised for highly mixed TeV-scale
top-squarks
\cite{Baer:2012up}.
As the name says, the higgsino mass parameter governs the masses of the higgsinos, 
and the requirement on $\mu$ translates into a requirement for the physical higgsino masses. 
This definition of naturalness restricts only the higgsino masses to be around 100-300\,GeV -- 
the remainder of the SUSY spectrum may easily lie in the multi-TeV range at little cost to naturalness.

In radiatively-driven natural SUSY models, the light higgsinos are almost mass degenerate with 
$\Delta m(\text{higgsinos})<20$ GeV. 
This is a challenging regime for the LHC although both CMS and ATLAS collaborations 
have now begun to explore this type of scenario via a soft $\ell^+\ell^- +jet+MET$ signature
arising from higgsino pair production \cite{Baer:2014kya}.
Their current limits probe $\mu\sim 100$ GeV and terminate at mass differences below 8 GeV \cite{CMS:2017fij}. 
Thus, there is a prospect that the required light higgsinos could only be probed by a lepton collider.

This is the motivation for our study of light, nearly mass-degenerate higgsinos at a lepton collider. 
The choice of experiment is the International Linear Collider or ILC whose construction is under 
political consideration in Japan. 
It would run with polarised electron and positron beams at $\sqrt{s}=250-500$ GeV. 
It has great potential for SUSY precision measurements as the beam backgrounds and the 
physics processes themselves are simple and controllable.

Our goal is to determine the exact potential for measuring higgsinos at the ILC. 
Furthermore, we want to test the capacity for exploiting the precision measurements 
to gain information on the unobserved, higher mass SUSY particles. 
Let us first discuss the precision measurements at the ILC.

\section{Prospects for measuring higgsinos at ILC}

We have studied three benchmark scenarios labelled as ILC1, ILC2 \cite{Baer:2014yta} and nGMM1 \cite{Baer:2016hfa} 
in detail. 
The common properties are that the only particles accessible via $e^+e^-$ collisions at 500 GeV 
are the four higgsinos $\LSP$, $\neutralinotwo$ and $\charginoone$. 
Their mass differences are $\sim$ 20 GeV (ILC1), 10 GeV (ILC2) and 5\,GeV (nGMM1).

The higgsino production and decay were simulated in a \texttt{Geant4}-based \cite{Agostinelli:2002hh} 
model of the International Large Detector (ILD), which is one of the planned detector concepts for the ILC \cite{TDR}. 
The simulated events were created with \texttt{Whizard} 1.95 \cite{Whizard}, 
taking into account the beam spectrum and ISR. 
The hadronisation was done with \texttt{Pythia} 6.422 \cite{PYTHIA}. 
The beam polarisation was assumed to be 80\% for the electrons and 30\% 
for the positrons with opposite chirality, $\mathcal{P}=(\pm80\%,\mp 30\%)$. The centre-of-mass energy was 500 GeV.

The following processes were studied: 
neutralino pair production and decay $e^+e^-\to \LSP \neutralinotwo \to \LSP \LSP e^+e^- (\mu^+\mu^-)$ 
and chargino pair production and decay $e^+e^-\to \widetilde{\chi}_1^+ \widetilde{\chi}_1^- \to \LSP \LSP q\bar{q}'e \nu_e (\mu \nu_\mu)$. 
Both processes have sizeable cross sections of 8-87 fb depending on beam polarisation in the ILC1 benchmark.

\begin{figure}[htbp]
 \begin{minipage}{0.48\linewidth}
  \begin{center}
    \includegraphics[width=0.99\textwidth]{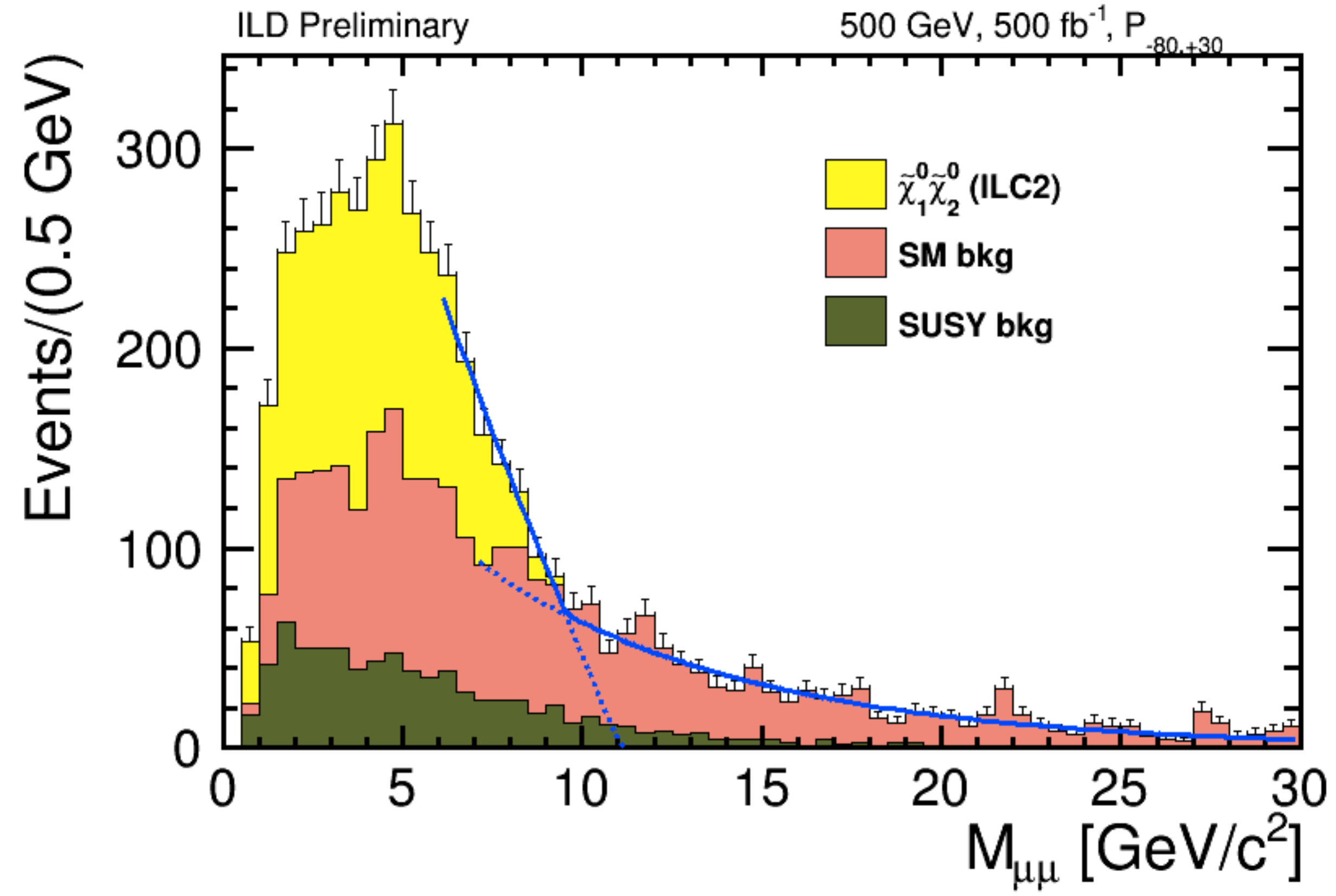}
  \end{center}
  \caption{The di-muon invariant mass in ILC2 neutralino decay.}
\label{fig:ILC2N1N2mass}
 \end{minipage}
 \hfill
 \begin{minipage}{0.48\linewidth}
  \begin{center}
    \includegraphics[width=0.94\textwidth]{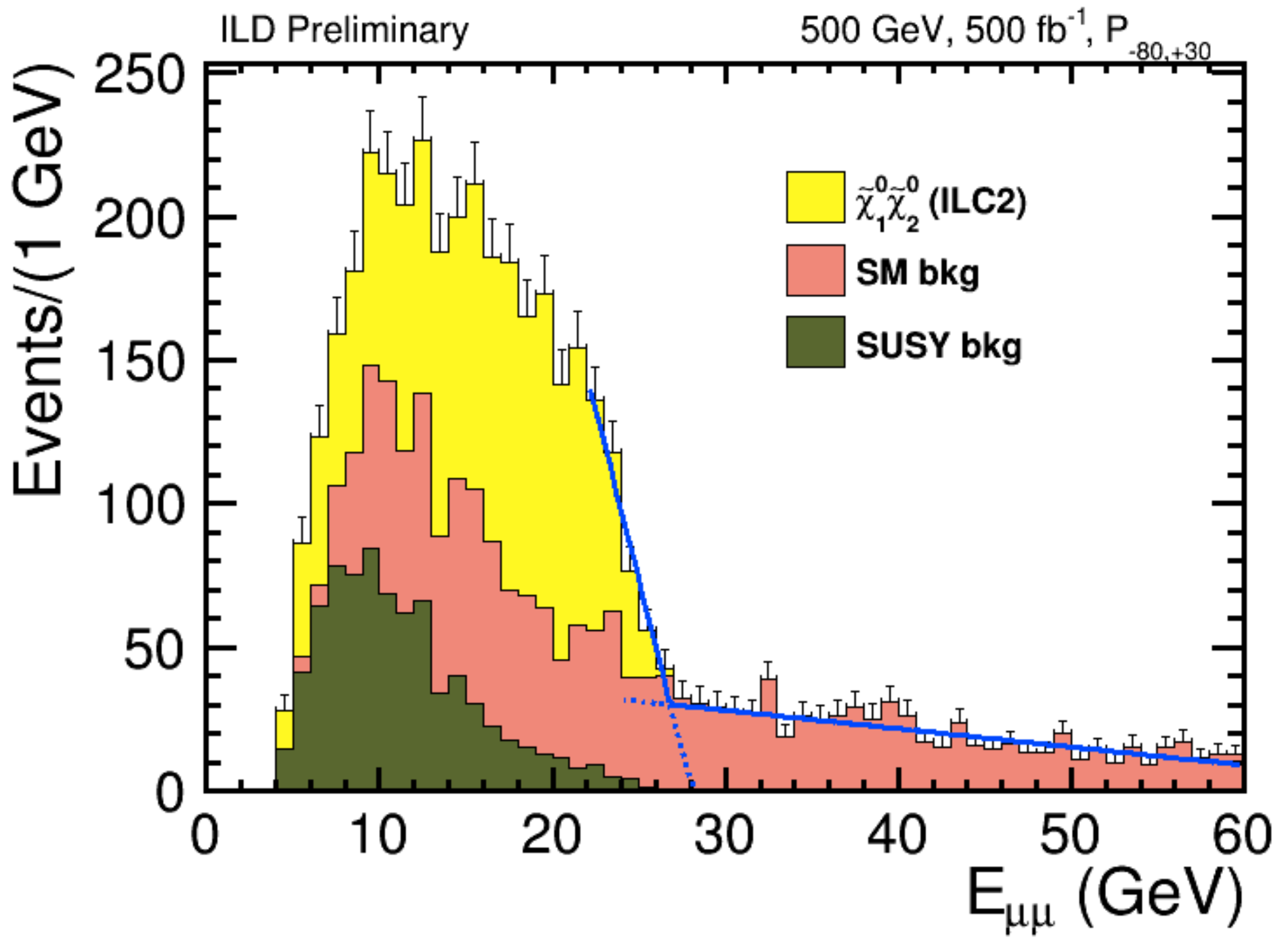}
      \end{center}
 \caption{The di-muon energy in ILC2 neutralino decay.}
 \label{fig:ILC2N1N2energy}
 \end{minipage}
\end{figure}

The higgsino massses can be measured relying on the zero-momentum and known centre-of-mass energy 
of the initial state. 
The invariant mass of the visible decay products of the neutralino or chargino can be reconstructed directly. 
Its maximum value corresponds to the mass difference of the neutralino or chargino with respect to the LSP. 
This and the maximum energy of the decay products can be used to solve for the absolute masses. 
The distributions of the energy and invariant mass of the higgsino decay products are plotted in 
Figs. \ref{fig:ILC2N1N2mass} and \ref{fig:ILC2N1N2energy} for ILC2. 
The prospects for the ILC1 benchmark are 0.2\% statistical uncertainty for higgsino masses 
with a total 4 ab$^{-1}$ data set, and sub-percent precisions are expected for ILC2 and nGMM1.

The higgsino cross sections can be measured by a cut and count strategy. 
For ILC1, the $\sigma \times$ BR of the studied channels are expected to be measured with 0.5-2\% statistical uncertainty 
depending on the channel and beam polarisation with a total 4 ab$^{-1}$ data set. 

\section{Results of SUSY parameter fits to measurements}

The higgsino measurements can be used to fit SUSY model parameters to gain further insights 
on yet unobserved, higher mass sparticles. 
Technically this was done via the fitting algorithm \texttt{Fittino} \cite{Bechtle:2004pc}, 
which uses Markov Chain Monte Carlo to probe the SUSY parameter space. 
As fit inputs, the higgsino masses and cross sections were used, 
along with the predicted Higgs mass and coupling measurements from the ILC \cite{Fujii:2015jha}. 
The SUSY spectrum was calculated with \texttt{SPheno} 3.3.9beta \cite{Porod:2003um} 
and the Higgs observables with \texttt{FeynHiggs2.10.2} \cite{Heinemeyer:1998yj}. 
A 10-parameter pMSSM model was fitted using $M_1$, $M_2$, $M_3$, $\tan \beta$, $\mu$, $m_A$, $M_Q(3)$, $M_U(3)$, a common sfermion mass M(sfermion) $=M_Q(1,2)=M_U(1,2)=M_D(1,2,3)=M_L(1,2,3)=M_E(1,2,3)$, and a common trilinear coupling $A_t=A_b=A_{\tau}$ as free parameters. 
Flavour mixing as well as trilinear coupling in the first and second generation was set to zero. 
This is a reasonable set of parameters for pMSSM as the underlying model for ILC1 and ILC2 is NUHM2 
and for nGMM1 mirage mediation. 

\begin{figure}[htbp]
 \begin{minipage}[t]{0.48\linewidth}
  \begin{center}
 \includegraphics[width=0.99\textwidth]{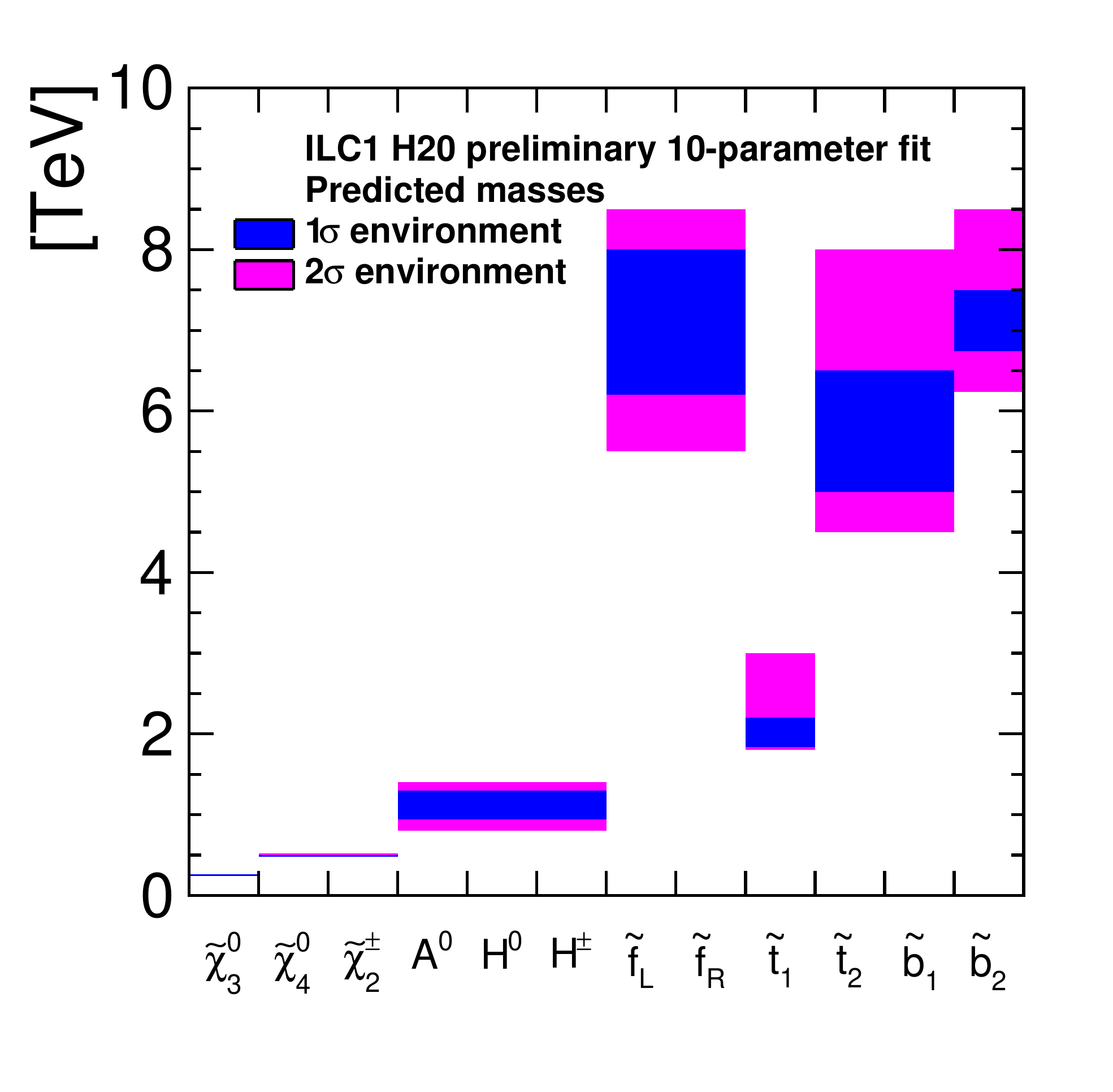}
  \end{center}
\caption{Predicted $1\sigma$ and $2\sigma$ ranges for the unobserved sparticles' masses in ILC1.}
\label{fig:ILC1predictedmasses}
 \end{minipage}
 \hfill
 \begin{minipage}[t]{0.48\linewidth}
  \begin{center}
    \includegraphics[width=0.99\textwidth]{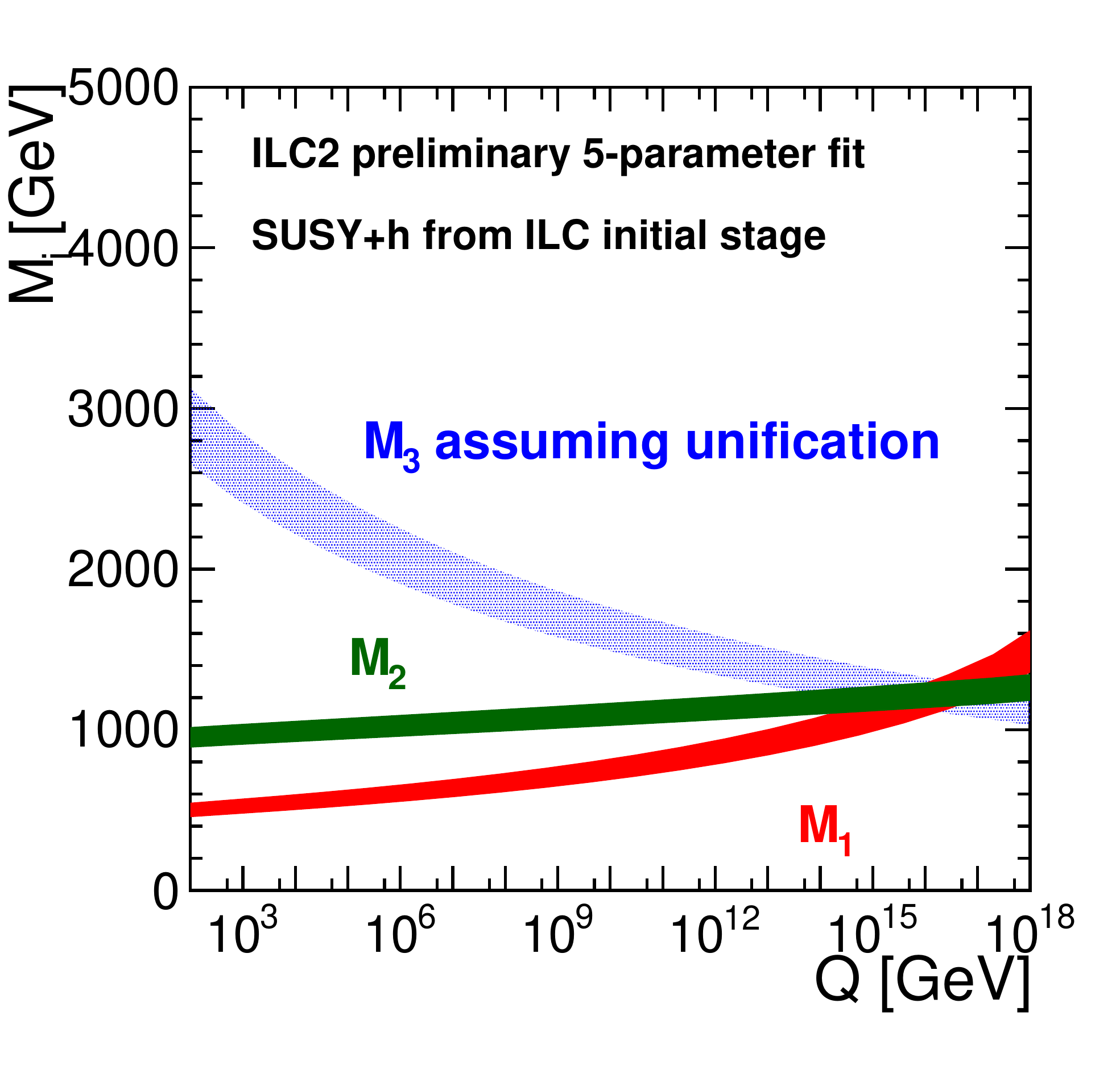}
          \end{center}
 \caption{The running of gaugino masses in ILC2. $M_1$ and $M_2$ are fitted. Their running gives the unification energy scale. If $M_3$ is assumed to unify too, then $M_3$ can be evolved down in energy.}
\label{fig:ILC2runningplot}
 \end{minipage}
\end{figure}

The result is that many parameters are determined or at least constrained. 
In particular, the bino and wino mass parameters which enter the higgsino mass splittings 
are determined with 1-2\% uncertainties. 
This means that the masses of the unobserved heavier neutralinos and charginos can be predicted 
with a 2\% uncertainty. 
Additionally, some 1$\sigma$ and 2$\sigma$ confidence intervals can be predicted for all other unobserved sparticles, 
as can be seen in Fig. \ref{fig:ILC1predictedmasses}. 
This is very exciting from the point of view of planning future high-energy colliders. 
The predictions of the bino and wino masses from the higgsino observations would give a clear guidance 
for the centre-of-mass energy of an ILC or LHC upgrade.

As an aside, since one of the higgsinos is the LSP, 
it is interesting to study the dark matter aspect of the benchmarks. 
A thermally-produced higgsino LSP provides only 5\% of the observed dark matter relic density. 
This feature can be predicted from the SUSY parameter fit via \texttt{MicrOMEGAs} \cite{Belanger:2001fz} 
very accurately. 
Should the higgsinos be observed, we would conclude either that the relic density requires additional 
particles such as axions or that the higgsino-like WIMP is produced non-thermally, perhaps via moduli decay.

As the parameters were fitted at the 1 TeV scale, further information may be obtained from their running. 
We ran the fitted parameters to the GUT scale using \texttt{SPheno}. 
The resulting 1$\sigma$ bands for $M_1$ and $M_2$ are plotted in Fig. \ref{fig:ILC2runningplot}. 
It can be seen that they unify at $2\times 10^{16}$ GeV. 
This is a hint of some SUSY GUT model underlying the measurements. 
If $M_3$ is assumed to unify at the same scale, then the running down in energy of $M_3$ 
is fixed and the gluino mass can be predicted.
Alternatively, if $M_1$ and $M_2$ unify at some intermediate scale, then evidence for mirage mediation may be gained.

\section{Conclusion}

We should expect light higgsinos from SUSY if naturalness arguments about the $\mu$ parameter hold. 
We have shown that the ILC would provide excellent prospects for both SUSY discovery and 
precision sparticle mass measurements if light higgsinos are within its kinematic range. 
The higgsino masses could be measured with percent-level precisions, 
while the higgsino cross sections could be measured with a few percent uncertainty. 
The precision measurements together with Higgs measurements constrain the unobserved SUSY spectrum, 
as was shown via a 10-parameter pMSSM fit. 
Furthermore, information about high-scale physics and grand unification 
can be obtained via running of the measured SUSY parameters. 

\section*{Acknowledgements}

We thank the ILD group for their work on event generation and simulation software development.

\end{document}